# Polarization Instability in Polydomain Ferroelectric Epitaxial Thin Films and the Formation of Heterophase Structures

N. A. Pertsev and V. G. Koukhar

*A. F. Ioffe Physico-Technical Institute, Russian Academy of Sciences,*

*194021 St. Petersburg, Russia*

A thermodynamic theory is developed for dense laminar domain structures in epitaxial ferroelectric films. It is found that, at some critical misfit strain between the film and substrate, the 90° *c/a/c/a* domain structure becomes unstable with respect to the appearance of the polarization component parallel to domain walls, which result in the formation of a heterophase structure. For PbTiO$_3$ and BaTiO$_3$ films, the stability ranges of polydomain and heterophase states are determined using "misfit strain-temperature" diagrams. Dielectric anomalies accompanying misfit-strain-driven structural transformations are described.

Ferroelectric thin films are intensively studied nowadays because their properties were found to be promising for microelectronic and micromechanical applications [1,2]. The formation of ferroelastic domain structures represents one of the basic mechanisms for reducing the strain energy appearing in epitaxial ferroelectric heterostructures due to a lattice mismatch between the film and substrate. Experimentally, polydomain states have been observed in PbTiO$_3$, Pb(Zr$_x$Ti$_{1-x}$)O$_3$, (Pb$_{1-x}$La$_x$)TiO$_3$, and KNbO$_3$ epifilms grown on various single crystalline substrates [3-11]. Numerous articles were devoted to the theoretical studies of the statics of elastic domains (twins) in epitaxial ferroelectric and ferroelastic films [12-14]. All these studies, however, were in fact performed in a linear elastic approximation, which neglects deviations of the order parameters (polarization components) in thin films from their equilibrium values in bulk crystals. At the same time, the Landau-Ginsburg-Devonshire-type thermodynamic theory, which was developed recently for single-domain ferroelectric thin films [15], shows that the mechanical film/substrate interaction may change the order parameters drastically. Therefore, a rigorous theoretical analysis is required to describe polydomain states in epitaxial heterostructures correctly.

In this Letter, a phenomenological thermodynamic theory is developed, which enables the determination of actual thermodynamic states inside dissimilar domains formed in epitaxial ferroelectric thin films. The theory is used to calculate polarization distributions in polydomain PbTiO$_3$ and BaTiO$_3$ films grown on cubic substrates and to describe their dielectric properties. To make the problem tractable mathematically, the polarization and strain fields are assumed to be homogeneous within each domain, which is a good approximation for "dense" laminar structures with domain widths much smaller than the film thickness.

Consider a polydomain single crystalline perovskite film epitaxially grown on a thick (001)-oriented cubic substrate. In accordance with the observations [3,4,7] the characteristic ferroelastic domain structure of an epitaxial film may be modeled by a periodic array of parallel domain walls. The geometry of this laminar structure is defined by the domain-wall periodicity $D$ and the volume fraction $\phi$ of domains of the first (or second) kind in the film. If the period $D$ is much smaller than the film thickness $H$, the polarization and strain fields within each domain may be regarded as homogeneous when calculating the total free energy F of a film. Indeed, in this case inhomogeneous internal fields become concentrated in two thin layers ($h \sim D$) localized near the film/substrate interface [12] and the film free surface. The contribution of the surface layers to the energy F is about $D/H$ times smaller than that of the inner part of the film and so may be neglected at $D << H$ in the first approximation. Moreover, the contribution of the self-energies of domain walls can be also ignored on the same grounds. In this approximation, equilibrium values of polarization components $P_i$ ($i = 1,2,3$) and lattice strains $S_n$ ($n = 1,2,3,...,6$ in the Voigt matrix notation) inside domains of the first and second kind and their equilibrium volume fractions become independent of the domain period $D$ and film thickness $H$. The calculation of these parameters defining piecewise homogeneous internal fields in the inner region of a polydomain film represents the goal of the present theory. The computation of the equilibrium domain period $D^*$ remains beyond the scope of this theory because $D^*$ is governed by the competition of the overall self-energy of domain walls and the energy stored in the surface layers. However, the equilibrium period $D^*$ may be evaluated in the elastic approximation [14]. The calculations show that the inequality $D^* << H$ indeed holds for sufficiently thick films ($H >> 100$ nm in the case of PbTiO$_3$ films [14]).

Let us determine now the form of the free-energy function, which ensures the correct thermodynamic description of polydomain thin films. To that end, we must specify the mechanical and electric boundary conditions of the problem. We focus here on the case of dielectric measurements, when there are no external mechanical forces acting on the film/substrate system, but the film is kept under an external electric field $\mathbf{E}_0$ induced via the top and bottom electrodes deposited on the upper surfaces of the film and substrate, respectively. Since the work done by extraneous mechanical sources equals zero, the free energy F of the heterostructure can be calculated simply by integrating the Helmholtz free-energy density F over the volume of the film/substrate system. However, this density must be taken in a modified form $\tilde{F} = F - \sum_{i=1}^{3} E_i(\varepsilon_0 E_i + P_i)$ [16], where **E** is the *internal* electric field and $\varepsilon_0$ is the permittivity of the vacuum, because electrostatic potentials of the electrodes are kept fixed in our case but not their charges. For periodic domain structures, the minimization of the total free energy F reduces to that of the energy density $<\tilde{F}>$ averaged over the period of domain pattern and the film thickness. In our case of a dense structure, the mean density $<\tilde{F}>$ can be evaluated simply as $<\tilde{F}> = \phi \tilde{F}' + (1-\phi) \tilde{F}''$, where $\tilde{F}'$ and $\tilde{F}''$ are the characteristic energy densities inside domains of the first and second kind. The Helmholtz free-energy function F may be approximated by a six-degree polynomial in polarization components $P_i$ ($i = 1,2,3$) [17]. Since for perovskite ferroelectrics the Gibbs energy function G is defined better than F [18,19], we shall derive the Helmholtz free energy F via the inverse Legendre transformation of G. This procedure gives $F = G + \sum_{n=1}^{6} \sigma_n S_n$ so that for ferroelectrics with a cubic paraelectric phase after some mathematical manipulation we obtain

$$\tilde{F} = \alpha_1(P_1^2 + P_2^2 + P_3^2) + \alpha_{11}(P_1^4 + P_2^4 + P_3^4) + \alpha_{12}(P_1^2 P_2^2 + P_1^2 P_3^2 + P_2^2 P_3^2)$$
$$+ \alpha_{111}(P_1^6 + P_2^6 + P_3^6) + \alpha_{112}[P_1^4(P_2^2 + P_3^2) + P_2^4(P_1^2 + P_3^2) + P_3^4(P_1^2 + P_2^2)] + \alpha_{123} P_1^2 P_2^2 P_3^2$$
$$+ \frac{1}{2}s_{11}(\sigma_1^2 + \sigma_2^2 + \sigma_3^2) + s_{12}(\sigma_1\sigma_2 + \sigma_1\sigma_3 + \sigma_2\sigma_3) + \frac{1}{2}s_{44}(\sigma_4^2 + \sigma_5^2 + \sigma_6^2)$$
$$- \frac{1}{2}\varepsilon_0(E_1^2 + E_2^2 + E_3^2) - E_1 P_1 - E_2 P_2 - E_3 P_3, \qquad (1)$$

where $\sigma_n$ are the internal mechanical stresses in the film, $\alpha_1$, $\alpha_{ij}$, and $\alpha_{ijk}$ are the dielectric stiffness and higher-order stiffness coefficients at constant stress [18,19], and $s_{mn}$ are the elastic compliances at constant polarization.

The mechanical and electric boundary conditions of the problem may be used now to eliminate the stresses and internal electric fields from the final expression for the mean energy density $<\tilde{F}>$ in the film, which follows from Eq. (1). Since changes of the in-plane sizes and shape of the film during its cooling from the deposition temperature are governed by a much thicker cubic substrate, the mean in-plane strains $<S_1>$ and $<S_2>$ must be equal to the misfit strain $S_m$ in the heterostructure, whereas the average shear strain $<S_6>$ should be zero in the rectangular reference frame with the $x_3$ axis perpendicular to the film/substrate interface. (We recall that the misfit strain $S_m$, which is the main internal parameter of the epitaxial couple, defines the macroscopic in-plane deformations $S_1 = S_2$ imposed by the substrate on a thin film and counted from the unstrained prototypic state of the latter [15].) Using the above equalities together with the relations $S_n = -\partial G/\partial T_n$ [19] between lattice strains, polarization, and mechanical stresses, we obtain the first three conditions imposed on the average internal stresses $<\sigma_n>$. Another three conditions, namely $<\sigma_3> = <\sigma_4> = <\sigma_5> = 0$, follow from the absence of the stress components $\sigma_3$, $\sigma_4$, and $\sigma_5$ on the free surface of the film. In turn, the mean electric field $<\mathbf{E}>$ in the film must be equal to the external field $\mathbf{E}_0$ defined by the potential difference induced between the electrodes.

The above nine macroscopic conditions may be supplemented with the "microscopic" boundary conditions which should be fulfilled on domain walls and relate internal fields existing in the adjacent domains of the first and second kind. In the rotated coordinate system ($x_1'$, $x_2'$, $x_3'$) with the $x_3'$ axis orthogonal to the domain walls, the requirement of strain compatibility gives $S_{1'}' = S_{1'}''$, $S_{2'}' = S_{2'}''$, and $S_{6'}' = S_{6'}''$. The mechanical equilibrium of the polydomain layer implies that $\sigma_{3'}' = \sigma_{3'}''$, $\sigma_{4'}' = \sigma_{4'}''$, and $\sigma_{5'}' = \sigma_{5'}''$. Finally, the continuity of the tangential components of the internal electric field and the normal component of the electric displacement yields $E_{1'}' = E_{1'}''$, $E_{2'}' = E_{2'}''$, and $\varepsilon_0 E_{3'}' + P_{3'}' = \varepsilon_0 E_{3'}'' + P_{3'}''$.

The introduced eighteen relationships enable us to express internal stresses $\sigma_n'$, $\sigma_n''$ and electric fields $E_i'$, $E_i''$ in domains of the first and second kind in terms of polarization components $P_i'$, $P_i''$ and the relative domain population $\phi$. Accordingly, the average energy density $<\tilde{F}>$ may be reduced to a function of seven variables: $P_i'$, $P_i''$ ($i=1,2,3$), and $\phi$. Performing numerically the minimization of $<\tilde{F}>(P_i', P_i'', \phi)$, we can find the equilibrium polarizations in both domains and the equilibrium domain population $\phi^*$. Evidently, these parameters appear to be functions of the misfit strain $S_m$ in the film/substrate system, temperature $T$, and the applied electric field $\mathbf{E}_0$. Comparing then the minimum energy $<\tilde{F}>^*(S_m, T, \mathbf{E}_0 = 0)$ of the polydomain film with the energies of various possible single-domain states [15], it is also possible to determine the stability range of a polydomain state in the misfit strain - temperature phase diagram of an epitaxial film.

We have performed the above calculations for PbTiO$_3$ (PT) and BaTiO$_3$ (BT) films using the coefficients $\alpha_1$, $\alpha_{ij}$, and $\alpha_{ijk}$ of the free-energy expansion given in Refs. [18,19] and the elastic compliances $s_{ij}$ and electrostrictive constants $Q_{ij}$ listed in Ref. [15]. In accordance with the observations [3,7,10] the domain walls were assumed to be inclined at 45° to the film/substrate interface. For this domain-wall orientation, the anticipated simplest solution of the problem corresponds to the standard "head to tail" polarization configuration observed in bulk perovskite crystals [17]. In the absence of external electric fields, the spontaneous polarization $\mathbf{P_s}$ has the same magnitude in the domains of the first and second kind ($c$ and $a$ domains), where $\mathbf{P_s}$ is orthogonal and parallel to the interface, respectively ($P_1^a = -P_3^c \neq 0$, $P_2^a = P_3^a = P_1^c = P_2^c = 0$, when the $x_2$ axis is directed along the walls). Internal electric fields and the stress components $\sigma_3$, $\sigma_4$, $\sigma_5$, and $\sigma_6$ are absent in this $c/a/c/a$ structure. The equilibrium volume fraction of $c$ domains equals

$$\phi_c^* = 1 - \frac{(s_{11} - s_{12})(S_m - Q_{12} P_s^2)}{s_{11}(Q_{11} - Q_{12}) P_s^2}, \quad (2)$$

and, at $\phi_c = \phi_c^*$, the stresses $\sigma_1^a$ and $\sigma_1^c$ also vanish, whereas the stress component $\sigma_2$ acquires the same value of $\sigma_2^a = \sigma_2^c = (S_m - Q_{12} P_s^2)/s_{11}$ in both domains. The magnitude of spontaneous polarization depends here on the misfit strain and varies according to the relation

$$P_s^2 = -\frac{\alpha_{11}^*}{3\alpha_{111}} + \left(\frac{\alpha_{11}^{*2}}{9a_{111}^2} - \frac{\alpha_1^*}{3\alpha_{111}}\right)^{1/2}, \quad (3)$$

where $\alpha_1^* = \alpha_1 - (Q_{12}/s_{11}) S_m$ and $\alpha_{11}^* = \alpha_{11} + Q_{12}^2/2s_{11}$ represent the renormalized coefficients of the free-energy expansion.

Numerical calculations confirm that the simple $c/a/c/a$ configuration may be stable in epitaxial ferroelectric films, but they also show that the stability range $R_d$ of this domain pattern in the misfit strain - temperature phase diagram is very limited, especially in BT (see Fig. 1). An evident restriction on the existence of the $c/a/c/a$ structure results from the inequality $0 < \phi_c^*(S_m) < 1$ imposed on the equilibrium volume fraction of $c$ domains. Equation (2) shows that

$\phi_c^*$ becomes equal to unity at $S_m^0(T) = Q_{12} P_0^2(T)$, where $P_0$ is the spontaneous polarization of a free bulk crystal. The transition line $S_m^0(T)$ between the monodomain $c$ phase and the $c/a/c/a$ structure constitutes the left-hand boundary of $R_d$ in the ($S_m$, $T$)-phase diagram (Fig. 1).

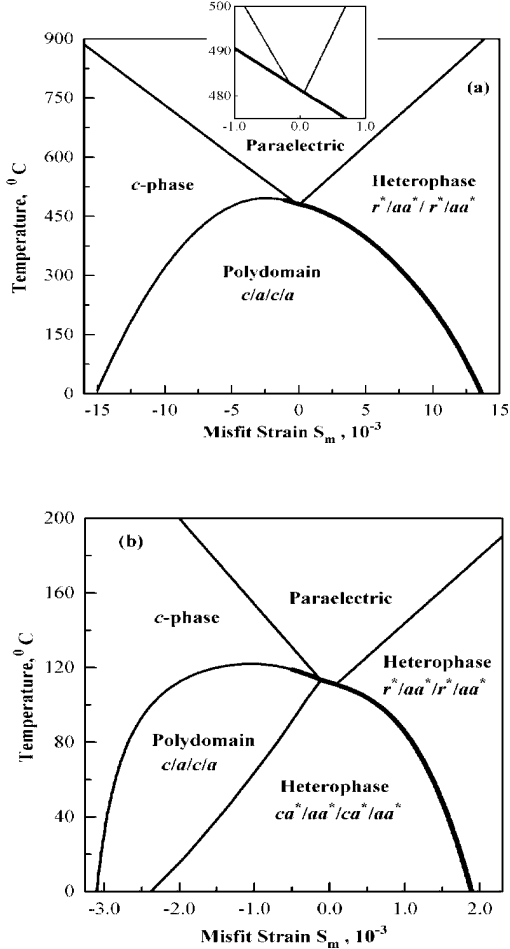

**Figure 1.** Phase diagrams of PbTiO$_3$ (a) and BaTiO$_3$ (b) epitaxial thin films grown on cubic substrates. The second- and first-order phase transitions are shown by thin and thick lines, respectively.

The curve $S_m^0(T)$ also defines the high-temperature part of the right-hand boundary, which adjoins the upper point of $R_d$ located at temperature $T_m = \theta + 2\varepsilon_0 C \alpha_{11}^2 / 3\alpha_{111}$ ($\theta$ and $C$ are the Curie-Weiss temperature and constant of the bulk crystal). This part of $S_m^0(T)$ relates to the second possible solution for $P_0^2$ that exists at $T > \theta$ in crystals with $\alpha_{11} < 0$ and $\alpha_{111} > 0$.

In PT, the next section of the right boundary is formed by a short segment of the straight line above which the solution (3) for the polarization in the $c/a/c/a$ state looses its physical meaning ( above this line $\alpha_{11}^{*2} < 3\alpha_1^* \alpha_{111}$ ). At the part of this segment situated between two triple points (see Fig. 1), the direct transformation of the paraelectric phase into the $c/a/c/a$ domain pattern takes place. The ferroelectric phase transition in this "misfit-strain window" near $S_m = 0$ is of the *first* order ($\alpha_{11}^* < 0$ in PT and BT), in constrast to the rest of the upper transition line, where it is of the *second* order [15].

The low-temperature part of the right boundary of $R_d$, surprisingly, was found to be situated at misfit strains $S_m^*(T)$ considerable smaller than the strains at which the equilibrium volume fraction of $c$ domains $\phi_c^*$ turns to zero. The detailed analysis shows that at $S_m > S_m^*$ the $c/a/c/a$ structure looses its stability against the appearance of the polarization component $P_2$ *parallel* to domain walls and the film/substrate interface. This results in the transformation of the polydomain state into a *heterophase* one. In PT, in both alternating phase layers all three components of the spontaneous polarization become different from zero (Fig. 2).

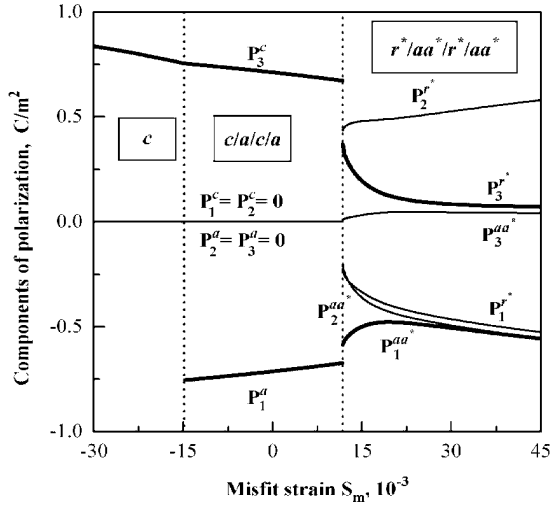

**Figure 2.** Polarization components in PbTiO$_3$ films as functions of the misfit strain $S_m$ at room temperature.

The transformed *a* domain, however, is distinguished by a very small out-of-plane polarization $P_3 << P_1, P_2$ at all misfit strains $S_m > S_m^*$. Moreover, the in-plane components $P_1$ and $P_2$ here acquire almost the same values at large misfit strains so that the forming phase may be regarded as a distorted *aa* phase [15]. In contrast, all three components of $\mathbf{P}_s$ have the same order of magnitude in the transformed *c* domain at $S_m \sim S_m^*$. Therefore, the second phase in the film should be considered as a distorted *r* phase [15] and the heterophase state may be termed *r\*/aa\*/r\*/aa\** structure. (The orthorhombic *aa* phase appears in monovariant epitaxial films at large positive misfit strains, whereas the monoclinic *r* phase forms in the gap between the stability ranges of *c* and *aa* phases[15].)

At the transition line $S_m^*(T)$ between the *c/a/c/a* and *r\*/aa\*/r\*/aa\** structures, the polydomain and heterophase states become energetically equivalent. The structural transformation here, according to abrupt changes of the polarization components at $S_m^*$ (see Fig. 2), corresponds to the first-order phase transition. In PT, at all $S_m > S_m^*$ the *r\*/aa\*/r\*/aa\** structure is energetically favored over monophase states and the calculations do not reveal the existence of a right boundary of the heterophase region $R_h$ in the diagram shown in Fig. 1a.

Strictly speaking, the stability of the *r\*/aa\*/r\*/aa\** array requires further analysis because laminar structures with internal boundaries orthogonal to the film/substrate interface are also possible in perovskite films [7,10]. Nevertheless, the discussed heterophase state must be at least metastable and could be observed by preparing a film with the *c/a/c/a* domain pattern first and then bending the substrate to increase the misfit strain above the critical value $S_m^*$. In this case appearing phase boundaries will retain the original orientation of 90° walls due to the stabilizing effect caused by their interaction with the underlying crystal lattice. The nature of the lattice orientational potential is similar to that of the Peierls potential relief hindering translations of domain walls in crystals [20]. The orientational relief, however, provides much higher barriers preventing rotations of internal boundaries away from the energetically favorable orientations along the {101} planes of the prototypic cubic phase. Since internal electric fields in the $r^*/aa^*/r^*/aa^*$ structure are relatively low (several kV/m), the discussed potential is expected to be sufficient to stabilize the predicted heterophase state.

Finally, we consider the dielectric properties of epitaxial ferroelectric films. The developed thermodynamic theory enables us to calculate numerically the *total* dielectric response of a polydomain or heterophase film, which results from both *intrinsic* and *extrinsic* contributions existing in this case. Indeed, the computations of equilibrium polarizations in the dissimilar domains or phase layers and their equilibrium volume fractions may be also performed for a given nonzero external electric field $\mathbf{E}_0$. Determining the dependence of the average polarization $<P_i> = \phi^* P_i' + (1-\phi^*) P_i''$ on $E_{0j}$, we can easily find the film permittivity $\varepsilon_{ij}$. Since the equilibrium volume fraction $\phi^*$ here is allowed to vary under field $\mathbf{E}_0$, the calculated dielectric constants $\varepsilon_{ij}$ involve an extrinsic contribution, which is caused by the field-induced displacements of domain walls (phase boundaries) from their equilibrium positions at $\mathbf{E}_o = 0$ [21]. These constants also correctly take into account the influence of the mechanical interaction with the substrate on the intrinsic (volume) contributions in a polydomain (heterophase) film.

Using the above approach, we computed the small-signal dielectric responses $\varepsilon_{ij}(\mathbf{E}_0 \to 0)$ of PT and BT films. Figure 3 shows variations of the diagonal components of the dielectric tensor $\varepsilon_{ij}$ with the misfit strain $S_m$ in PT films. It can be seen that the transformation of $c$ phase into $c/a/c/a$ pattern, which should be considered as a structural transition between two polar states, manifests itself in jumps of $\varepsilon_{11}$ and $\varepsilon_{33}$ at the transition line $S_m^0(T)$. In contrast, strong dielectric anomalies accompany the transition between polydomain and heterophase states. Remarkably, the maxima of $\varepsilon_{11}$ and $\varepsilon_{22}$ are shifted from the transition line $S_m^*(T)$, whereas the peak of $\varepsilon_{33}$ is located just at the critical strain $S_m^*$ (Fig. 3).

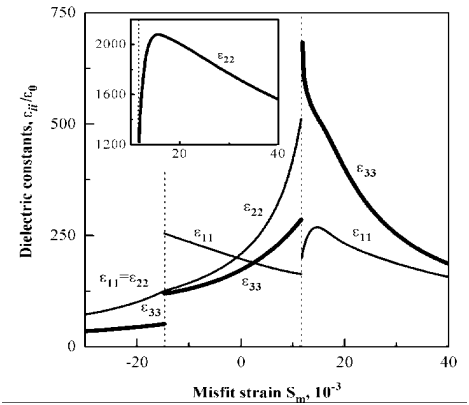

**Figure 3.** Dependences of the dielectric constants $\varepsilon_{ii}$ ($i = 1,2,3$) of PbTiO$_3$ films on the misfit strain $S_m$ at room temperature.